\documentclass{elsart}

\usepackage{graphicx}
\usepackage{epsfig}

\usepackage{amssymb}
\usepackage{eucal}

\newcommand{\gev}{$\mathrm{(GeV/c)}^{2}$}
\newcommand{\qsq}{$\mathrm{Q^2}$}
\newcommand{\qvec}{$\mathrm{\vec{q}}$}
\newcommand{\eep}{$(e,e'p)$}   
\newcommand{\een}{$(e,e'n)$}   
\newcommand{\he}{$\mathrm{^3He}$}
\newcommand{\vhe}{$\mathrm{^3\overrightarrow{\mathrm{He}}}$}

\newcommand{\heen}{$\mathrm{^3\overrightarrow{\mathrm{He}}(\vec{e},e'n)}$}

\newcommand{\hneen}{$\mathrm{^3\overrightarrow{\mathrm{He}}(e,e'n)}$}
\newcommand{\hneep}{$\mathrm{^3\overrightarrow{\mathrm{He}}(e,e'p)}$}
\newcommand{\gen}{$G_{en}$}
\newcommand{\gmn}{$G_{mn}$}
\newcommand{\gep}{$G_{ep}$}
\newcommand{\gmp}{$G_{mp}$}

\newcommand{\ape}{$A_{\perp}$}
\newcommand{\apa}{$A_{\parallel}$}
\newcommand{\ayz}{$A_{y}^o$}
\newcommand{\et}{{\em et al.}}

\begin{document}

\begin{frontmatter}
%
\title{The neutron charge form factor and target analyzing powers from 
\mbox{$\mathrm{^3\vec{He}(\vec{e},e'n)}$}--scattering}


\author[g]{J.~Bermuth}
\author[f]{P.~Merle}
\author[a]{C.~Carasco}
\author[f]{D.~Baumann}
\author[f]{R.~B{\"o}hm}
\author[k]{D.~Bosnar}
\author[f]{M.~Ding}
\author[f]{M.O.~Distler}
\author[f]{J.~Friedrich}
\author[f]{J.M.~Friedrich}
\author[d]{J.~Golak}
\author[b]{W.~Gl{\"o}ckle}
\author[a]{M.~Hauger}
\author[g]{W.~Heil} 
\author[f]{P.~Jennewein}
\author[a]{J.~Jourdan\corauthref{cor1}} 
\corauth[cor1]{Corresponding author, e-mail: Juerg.Jourdan@unibas.ch}
\author[i]{H.~Kamada}
\author[h]{A.~Klein}
\author[c]{M.~Kohl}
\author[a]{B.~Krusche} 
\author[f]{K.W.~Krygier} 
\author[f]{H.~Merkel}
\author[f]{U.~M\"uller}
\author[f]{R.~Neuhausen}
\author[j]{A.~Nogga}
\author[a]{Ch.~Normand}
\author[g]{E.~Otten}
\author[f]{Th.~Pospischil}
\author[e]{M.~Potokar}
\author[a]{D.~Rohe}
\author[f]{H.~Schmieden}
\author[g]{J.~Schmiedeskamp}
\author[f]{M.~Seimetz}
\author[a]{I.~Sick}
\author[e]{S.~\v{S}irca}
\author[d]{R.~Skibi{\'n}ski}
\author[a]{G.~Testa}
\author[f]{Th.~Walcher}
\author[a]{G.~Warren}
\author[f]{M.~Weis}
\author[d]{H.~Wita{\l}a}
\author[a]{H.~W{\"o}hrle}
\author[a]{M.~Zeier}

\address[a]{Dept. f\"ur Physik und Astronomie, Universit\"at Basel, Switzerland}
\address[b]{Insitut f\"ur Theoretische Physik II, Ruhr--Universit\"at Bochum, Germany}
\address[c]{Institut f\"ur Kernphysik, Technische Universit\"at Darmstadt, Germany}
\address[d]{Institute of Physics, Jagiellonian University, Krak{\'o}w, Poland}
\address[e]{Institute Jo\v{z}ef Stefan, University of Ljubljana, Ljubljana, Slovenia}
\address[f]{Insitut f\"ur Kernphysik, Johannes Gutenberg--Universit\"at, Mainz, Germany}
\address[g]{Insitut f\"ur Physik, Johannes Gutenberg--Universit\"at, Mainz, Germany}
\address[h]{Dept. of Physics, Old Dominian University, Norfolk, USA}
\address[i]{Dept. of Physics, Kyushu Institute of Technology, 
Tobata, Kitakyushu, Japan}
\address[j]{Dept. of Physics, University of Arizona, Tucson, Arizona, USA}
\address[k]{Dept. of Physics, University of Zagreb, Croatia}

\begin{abstract}
The charge form factor of the neutron has been determined from
asymmetries measured in quasi--elastic \heen\ at a momentum 
transfer of 0.67~\gev. In addition, the target analyzing power, \ayz,
has been measured to study effects of final state interactions and meson
exchange currents.
\end{abstract}

\vspace*{-4mm}

\end{frontmatter}


{\bf Introduction:} \hspace*{0.01cm}
\label{intro}
The form factors of the nucleon are fundamental observables. Precise data
allow for sensitive tests of the theory of the strong 
interaction --- quantum chromodynamics (QCD) --- in the non-perturbative regime. 
A complete study of the theoretical concepts requires measurements 
not only for the proton but also for the neutron. Accurate data at low momentum
transfer are also required for the calculation of nuclear form factors.

Due to the lack of a free neutron target only neutrons bound in light nuclei can be 
studied. In this case, determinations of the charge, \gen, and magnetic, \gmn, form factor
from elastic or quasi--elastic cross section data via the Rosenbluth technique do not 
lead to data with the desired precision. The subtraction of the proton
contribution, theoretical corrections due to the unfolding of the nuclear structure
and corrections to final state interaction (FSI) and meson--exchange currents (MEC) 
limit the accuracy to $\sim$30\%. 

Measurements of precise data of the neutron form factors became possible 
by means of alternative techniques exploiting polarized electron beams and 
polarized targets or recoil polarimeters. The technique to determine \gen\ with
a precision of $<$10\% relies on asymmetry measurements in quasi--free $(e,e'n)$ 
coincidence experiments in which the asymmetry is given by the interference term 
and is proportional to $G_{en}\cdot G_{mn}$ in the Plane Wave Impulse Approximation (PWIA). 
The small contribution of \gen\ is amplified 
by the large value of \gmn\ and a measurement of the asymmetry allows for
significant improvements in the precision \cite{Arnold81,Blankleider84,Arenhoevel88}.
The continuous wave (cw) electron beams available today allowed for the determination of \gmn\
with accuracies of $\sim$2\% \cite{Kubon01}.

Because of its low binding energy, the deuteron is usually employed for studies of 
neutron properties. However, for polarization studies \he\ is particularly suitable due to the 
fact that for the major part of the ground state wave function the spins of the two 
protons are coupled antiparallel, so that spin dependent observables are dominated 
by the neutron \cite{Golak01}. In addition, at least at low \qsq, corrections due to
nuclear structure effects, FSI, and MEC can be calculated using modern three-body 
calculations. These 
calculations allow for a quantitative description of the three-nucleon system with 
similar precision as for the deuteron \cite{Kievsky98,Gloeckle99}. 

The asymmetry in double polarization experiments is determined with
\begin{equation}
A(\theta^*,\phi^*) = \frac{1}{P_e \cdot P_t} \cdot \frac{N^+ - N^-}{N^+ + N^-}
\label{eq_asym}
\end{equation}
where $\theta^*,\phi^*$ are the polar and the azimuthal angle of the target spin 
direction with respect to the three momentum transfer \qvec. The polarizations of beam 
and target are given by $P_e$ and $P_t$ and the normalized \heen\ events 
for positive (negative) electron helicity are $N^+ (N^-)$. With the target 
spin orientation parallel and perpendicular to \qvec\ two independent asymmetries 
$A_{\parallel} = A(0^o,0^o)$ and $A_{\perp} = A(90^o,0^o)$ can be measured. 
In PWIA \gen\ can then be determined via
\begin{equation}
G_{en}^{PWIA} = \frac{b}{a} \cdot G_{mn} \frac{(P_e P_t V)_{\parallel}}{(P_e P_t V)_{\perp}} 
\frac{A_{\perp}}{A_{\parallel}}
\label{eq:gen}
\end{equation}
with the kinematical factors $a$ and $b$ \cite{Raskin89}. The factor  $V$  accounts 
for a possible dilution due to contributions with vanishing asymmetry. As $P_e$, $P_t$,
and $V$ do not depend on the target spin orientation they cancel in principle in the determination
of $G_{en}^{PWIA}$. In practice, \apa\ and \ape\ are measured in sequence, as such $P_e$
and/or $P_t$ may change during the two asymmetry measurements. It will be discussed 
below that such changes are measured and accounted for.

In order to study the FSI--corrections necessary for the 
determination of \gen\ the target analyzing power \ayz\ provides an experimental quantity 
that is sensitive to these effects. 
For an unpolarized beam and the target spin aligned perpendicular to the scattering plane 
the target analyzing power can be measured with
\begin{equation}
A_{y}^o = \frac{1}{P_t} \cdot \frac{N^{\uparrow} - N^{\downarrow}}{N^{\uparrow} + N^{\downarrow}}
\end{equation}
where $N^{\uparrow},(N^{\downarrow})$ are the normalized 
$\mathrm{^3\overrightarrow{\mathrm{He}}(\vec{e},e'N)}$ events for target spin 
aligned parallel (antiparallel) 
to the normal of the scattering plane. For coplanar scattering \ayz\ is identical to zero in PWIA due 
to the combination of time reversal 
invariance and hermiticity of the transition matrix \cite{Conzett98}. Thus, a non--zero value of 
\ayz\ signals FSI and MEC effects and its measurement provides a check of the calculation 
of these effects.

The present paper reports about a new determination of \gen\ from measurements 
of \ape\ and \apa\ of \heen--scattering at a four momentum transfer of \qsq=0.67\gev. 
The same kinematics  is chosen as for the measurements by Rohe \et\ \cite{Rohe99}.
In addition, the same technique and almost the same apparatus is employed which allows
to combine the data reducing the statistical error bar of \gen\ by almost a factor 
of two. 

Consequently, the improved precision requires a careful determination of FSI and MEC effects.
Target analyzing powers \ayz\ have been measured at the same \qsq\ and at \qsq=0.37\gev\ in 
order to properly determine FSI and MEC corrections of the combined result.

{\bf Experimental Setup:} \hspace*{0.01cm}
\label{exp}
At the Mainz Microtron (MAMI) \cite{Herminghaus90} longitudinally polarized electrons 
with a  polarization of $\sim$0.8 were produced with a strained layer GaAsP crystal at a typical 
current of $10~\mu A$ \cite{Aulenbacher97}. The polarized cw electron beam was accelerated
to an energy of 854.5~MeV and guided to the three--spectrometer hall  \cite{Blomqvist97}.
The \vhe--target consisted of a spherical glass container with two cylindrical extentions 
sealed with oxygen--free 25$\mu$m Cu--windows. Coating the glass container with Cs led
to relaxation times of about 80h. The Cu--windows were positioned outside of the 
acceptance of the spectrometer ($\sim$5~cm) and shielded with Pb--blocks to minimize background
from beam--window interactions. The $^3{\rm He}$--target was polarized via metastable optical pumping
to a typical polarization of 0.5 and compressed to an operating pressure of 4~bar
 \cite{Surkau97}. 

Spectrometer A with a solid angle of 28~msr and a momentum acceptance of 20\% 
was used to detect the quasi--elastically scattered electrons at a scattering angle of 78.6$^o$. 
The recoiling nucleons were detected in coincidence with an array of plastic scintillator bars 
placed at 32.2$^o$, the direction of \qvec\ for the maximum of the quasi--elastic peak.

The hadron detector consisted of an array of four layers of five 
plastic scintillator bars with dimensions $50\times10\times10$~cm$^3$
preceeded by two 1~cm thick $\Delta$E detectors for particle identification. 
The detector was placed at a distance  of 160~cm from the 
target, resulting in a solid angle of 100~msr. The neutron efficiency during
the present experiment was determined to 18.3\%. The entire detector
was shielded with 10~cm Pb except for an opening towards the
target were the Pb--shield was reduced to 2~cm. 

The entire \vhe--target was enclosed in a rectangular box of 
2~mm thick $\mu$-metal and iron. The box served as an effective shield for the 
stray field of the  magnetic  spectrometers 
and provided a homogeneous magnetic guiding field of $\approx 4 \cdot 10^{-4}$~T 
produced by three independent pairs of coils. With additional correction coils a relative field 
gradient of less than $5 \cdot 10^{-4}$\,cm$^{-1}$ was achieved. 
The setup also allowed for an independent rotation of the target spin
in any desired direction with an accuracy of 0.1$^o$ by remote control. 

The product of target and beam
polarization was monitored during the data taking via determination of
the asymmetry for elastic \vhe$(\vec{e},e)$--scattering for which the form factors,
hence the asymmetries, are accurately known \cite{Amroun94}. The analysis of these 
data resulted in a polarization product of $0.279\pm 0.010$ for runs with 
$A = |A_{\parallel}|$  and $0.282\pm 0.003$ for $A = |A_{\perp}|$. 
The different error bar results from the sensitivity of elastic scattering
to the target spin direction. 

In addition, the time dependence of the polarization of the target cell was continuously 
measured during the experiment by Nuclear Magnetic Resonance, while the absolute 
polarization was measured by the method of Adiabatic Fast Passage \cite{Wilms97}.
The mean target polarization from these measurements was $0.356\pm 0.015$. 
>From the elastic scattering data and the target polarization 
measurements a beam polarization of $P_e = 0.788\pm 0.036$ was extracted which agreed
well with the determination with a M{\o}ller polarimeter ($0.827\pm 0.017$).

{\bf Determination of \gen:} \hspace*{0.01cm}
To determine \gen\ the asymmetries \ape\ and \apa\ of \heen\ have been 
measured. The same kinematics was chosen as in \cite{Rohe99} with the 
motivation to combine the two measurements hereby decreasing the statistical error bar
of \gen.

In the analysis of the data the neutron is identified with a cut on the coincidence time and 
the absence of a hit in the $\Delta$E--amplitude spectrum for two consecutive $\Delta$E--detectors. 
Neutrons from (p,n) charge exchange in the Pb-shielding contribute in first order to the dilution 
factor $V$, but the effect cancels in the determination of \gen\ through equation \ref{eq:gen}. 

In order to minimize the dependence on the target polarization,
data were accumulated alternatively for \ape\ and \apa\ at regular intervals 
by corresponding rotations of the target spin.
The polarization ratio that enters in the determination of \gen\ 
(see equation \ref {eq:gen}) was unity within 2.6\%.

Experimental corrections have been determined via Monte Carlo simulation of the experiment based on PWIA
with the momentum distribution compiled by Jans \et\ \cite{jans87}. Accounting for energy loss via bremsstrahlung and for 
asymmetric angle and momentum acceptances of the spectrometer and the hadron detector can be reliably done
\cite{Rohe99}. 
The dominant correction is due to the asymmetric acceptance of the electron spectrometer which leads to an
angle shift of \qvec. The resultant effect is an enhancement of the measured \ape--value due to the 
contribution proportional to $G_{mn}^2$. Bremsstrahlung and missing energy lead to a similar effect. 
The total correction from these effects amounts to $-7.4\pm$3.0\%.

Finally, the value for the magnetic form factor required for the determination of \gen\ is taken from
the parameterization by Kubon \et\  \cite{Kubon01} with \gmn= (1.037$\pm$0.012)$\cdot \mu_n G_D$
where $\mu_n$ is the magnetic moment of the neutron in units of nuclear magnetons and $G_D$ the dipole form factor.
The resulting experimental value is $G_{en}^{PWIA}=0.0416\pm 0.0102_{stat} \pm 0.0024_{syst}$.

This value is in good agreement with the value by Rohe \et\ \cite{Rohe99}. 
A weighted average of the two values leads to $G_{en}^{PWIA}=$0.0468 $\pm 0.0064_{stat}$ $\pm 0.0027_{syst}$ which
corresponds to a reduction of the statistical error bar by almost a factor of two.
	
{\bf Target analyzing power:} \hspace*{0.01cm}
\label{aynull}
The target analyzing power \ayz\ has been measured for \hneen\ and \hneep\ at \qsq=0.67\gev\ (the 
kinematics of the \gen--measurement) and at 0.37\gev. 
The measurement at 0.37\gev\ was performed by lowering the beam energy to 600~MeV as the geometrical 
constraints of the target shielding box and the hadron detector did not permit a
change of the scattering or recoil angle. An unpolarized beam was used and
the target spin was aligned perpendicular to the scattering plane and reversed every 2~minutes.

The analysis of the \ayz--data is very similar to the one described above. 
Electrons are accepted for energy transfers $\omega = 225-290$~MeV (314--408~MeV) for the 
low (high) \qsq--point. The hadron is identified with a cut on the coincidence 
time and the $\Delta$E--amplitude spectrum. 

Contrary to the determination of $G_{en}^{PWIA}$, dilution effects do not cancel for \ayz\
and have to be determined. The 2cm Pb--absorber of the hadron detector leads to misidentified 
proton/neutron events due to charge exchange scattering in the Pb--absorber.
The 3 times larger ${e-p}$ cross section and the 5 times larger efficiency of 
the hadron detector for protons leads to a dilution effect that is negligible for $A^o_{y(e,e'p)}$
but must be taken into account for $A^o_{y(e,e'n)}$. 

The correction was measured by replacing the \he--gas in the target with hydrogen and tagging the 
recoil protons with the elastically scattered electrons. The fraction of protons, misidentified 
as neutrons amounts to $0.18\pm0.01$ ($0.13\pm0.01$) for the low (high) \qsq--point.
An additional contribution results from the uncorrelated background in the coincidence time spectrum 
determined to 0.056(0.025) for the $(e,e'n)$--events.

The $A^o_{y(e,e'n)}$--values have been corrected according to 
\begin{equation}
A^o_{y(e,e'n)} =  \frac{(A^o_{y~total} - x A^o_{y~back})}{(1-x)}
\end{equation} 
with $x$ the total fraction of background events, $ A^o_{y~back}$ its analyzing power and
$A^o_{y~total}$ the analyzing power of the total $(e,e'n)$--yields.
\begin{table}[htb]
\begin{center}
\parbox{13cm}{\caption[]{
Results of \ayz\ for the \hneen\ and \hneep--reactions. The experimental data 
at \qsq=0.37\gev\ are compared to results of a complete Faddeev calculation. 
For \een\ the effects of dropping different contributions in the calculation are also shown.
}\label{ayz}} \\ [4mm]
\begin{tabular}{|ccccccc|l|r|}
\hline
 & & & & & & & & \\ [-5mm]
\multicolumn{7}{|l|}{\qsq\ $(GeV/c)^2$}& ~~~~~~0.37 & 0.67~~~~~~ \\ [1mm]
\hline
 & & & & & & & & \\ [-5mm]
\multicolumn{7}{|l|}{$\mathrm{^3\overrightarrow{\mathrm{He}}(e,e'n)}$:} & &  \\
\multicolumn{7}{|l|}{Experiment} &
0.144$\pm$0.034 & 0.028$\pm$0.010 \\
\multicolumn{7}{|l|}{Theory}  & 0.178 & \\
\multicolumn{7}{|l|}{Theory without MEC} & 0.186 & \\
\multicolumn{7}{|l|}{Theory with \gep=\gmp=0} & 
0.004 & \\ [1mm]
\hline
 & & & & & & & & \\ [-5mm]
\multicolumn{7}{|l|}{$\mathrm{^3\overrightarrow{\mathrm{He}}(e,e'p)}$:} & &  \\
\multicolumn{7}{|l|}{Experiment}  &
 --0.025$\pm$0.005 & --0.016$\pm$0.005 \\
\multicolumn{7}{|l|}{Theory} & --0.017 &
 \\ [1mm]
\hline 
\end{tabular}
\end{center}
\end{table}

The corrected experimental results for \ayz\ are shown in table \ref{ayz}. Total error bars
are given. The errors are dominated by statistics with a small contribution of systematic 
errors due to false asymmetries and polarization measurements.

For both \een\ and \eep\ the agreement of the experimental values at \qsq=0.37\gev\ with 
the result of a complete calculation by Golak \et\ \cite{Golak02} is quite satisfactory. 
Neglecting the contribution of MEC in the calculation  has little effect on \ayz\ for \een. 
On the other hand a calculation for \een\ was also performed setting the proton
form factors \gep\ and \gmp\ to zero. As can be seen from table \ref{ayz} the
effect is quite dramatic suggesting that 98\% of the FSI--effect measured with $A^o_{y(e,e'n)}$
results from a coupling of the virtual photon to the proton followed by 
a $(p,n)$ charge exchange reaction in the three--body--system. 

A similar theoretical study is not possible at \qsq=0.67\gev\ due to the non--relativistic
nature of present day calculations and the fact that the transferred energy 
is well above the pion production threshold.

{\bf FSI corrections of \gen:} \hspace*{0.01cm}
\label{fsigen}
For the same reason the FSI--effects in \ape/\apa\ which are needed as corrections to determine 
\gen\ can not be calculated at this \qsq\ using today's non--relativistic Faddeev codes. 
However, we will discuss in the following that
with the measurements of \ayz\ and the measurements by Carasco \et\ \cite{Carasco03}
a reliable estimate of the effects can be made. In this approach, we first determine $G_{en}^{PWIA}$
which accounts for relativistic kinematics --- the only significant relativistic effect to
consider at this \qsq\  \cite{Carasco03} --- and then apply the FSI corrections based on the 
acquired experimental information. 

Two effects contribute to the FSI correction and have to be considered at first order.
First, the photon couples to one of the protons followed by a charge exchange process in the
three--body--system simulating an \een--event. At \qsq=0.37\gev\ the complete 
Faddeev calculation by Golak \et\ \cite{Golak02} which successfully predicted
\ayz\ predicts a total FSI effect for the ratio \ape/\apa\ of 25\%. The calculations
also show that the charge exchange process which is responsible for 98\%
of \ayz\ amounts to 60\% of the total FSI--effect in \ape/\apa.

The ratio of the elementary cross sections $\sigma_{e-p}/\sigma_{e-n}$, which is a 
measure for the probability of the photon coupling to a proton or a neutron is similar 
at \qsq\ of 0.37\gev and 0.67\gev.  We therefore assume that the charge exchange 
process also contributes with 60\% to the FSI effect in \ape/\apa\ at 0.67\gev.

With the experimental knowledge of \ayz\ at both \qsq--values the contribution of the 
charge exchange processes in \ape/\apa\ can be determined with the ratio of the 
experimental \ayz--values scaling the effect to 0.67\gev.
This results in a FSI correction of 3\% in \gen. 
  
Second, the photon couples to the neutron followed by a rescattering process in the 
three--body--continuum which may also lead to FSI--effects. 
The effect for this type of FSI in \apa\ and \ape\ of \eep\ has been discussed 
in detail by Carasco \et\  \cite{Carasco03}. The results of a calculation which treats only the 
FSI between the two (slow) spectator nucleons agree well with the 
experimental data. The same calculation has been used to compute \apa\ and \ape\
of \een. Contrary to the significant FSI effect for \ape\ and
\apa\ of \eep\ observed in \cite{Carasco03} the corresponding contribution is small for 
the asymmetries of \een. 
The resulting FSI--effect of this process averaged over the accepted phase space is 0.4\% in \gen.

\begin{figure}[hbt]
\begin{center}
\includegraphics[scale=0.7,clip]{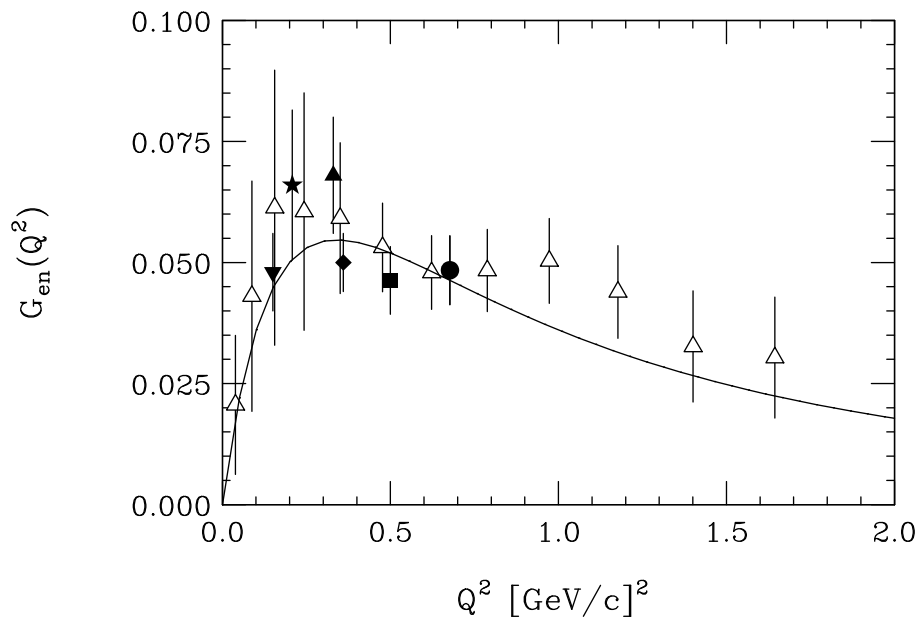}
\parbox{13cm}{\caption[]{
Experimental results of \gen. Shown are the results from double-polarization 
experiments, the present result {\protect{($\bullet$), \cite{Ahmidouch01}($\blacksquare$), 
\cite{Becker99}($\blacklozenge$), \cite{Herberg99}($\blacktriangledown$), \cite{Ostrick99}($\blacktriangle$),
and \cite{Passchier99}($\bigstar$),}} and the results from the elastic quadrupole form factor
{\protect{\cite{Schiavilla01},($\triangle$)}}. The solid line is the parameterization 
by Galster {\protect{\cite{Galster71}}}
}\label{gen}} 
\end{center} 
\end{figure}

Thus, we conclude that the total FSI--correction to \gen\ at 0.67\gev\ is small (of the order of
3.4\%) and dominated by charge--exchange processes. The correction is accounted for in the final
result with a relative uncertainty of 50\% added in 
quadrature to the quadratic sum of the experimental uncertainties of the combined \gen--result. 
The final value of \gen=$0.0484\pm0.0071$ is shown in figure \ref{gen}. This result is in excellent
agreement with the \gen--values deduced from 
the quadrupole form factor of elastic $e-d$--scattering \cite{Schiavilla01}. 

{\bf Conclusions:} \hspace*{0.01cm}
In the present experiment, \gen\ has been measured via the double polarization reaction \heen. It 
has greatly improved the accuracy of our knowledge of \gen\ from such measurements at \qsq=0.67\gev.
The applied contribution from FSI is estimated as (3.4$\pm$1.7)\% at this high \qsq\ which is considerably 
smaller than the statistical uncertainty. The good agreement of the final value of 
\gen=$0.0484\pm0.0071$ with data from other double polarization experiments corrected for FSI
is very satisfactory. The value for \gen\ also agrees well with \gen--values extracted 
from the quadrupole form factor determined in elastic $e-d$--scattering.

{\bf Acknowledgments:} \hspace*{0.01cm}
\label{ackn}
This work was supported by the Schweizerische Nationalfonds, Deutsche
Forschungsgemeinschaft (SFB 443), the Polish Committee for Scientific Research, 
the Foundation for Polish Science and
the NSF. The numerical calculations have
been performed on the Cray SV1 of the NIC, J\"ulich, Germany.



\end{document}